# Performance of ChatGPT on the Test of Understanding Graphs in Kinematics

Giulia Polverini and Bor Gregorcic

Department of Physics and Astronomy, Uppsala University, Box 516, 75120 Uppsala, Sweden

**Abstract:** The well-known artificial intelligence-based chatbot ChatGPT-4 has become able to process image data as input in October 2023. We investigated its performance on the Test of Understanding Graphs in Kinematics to inform the physics education community of the current potential of using ChatGPT in the education process, particularly on tasks that involve graphical interpretation. We found that ChatGPT, on average, performed similarly to students taking a high-school level physics course, but with important differences in the distribution of the correctness of its responses, as well as in terms of the displayed "reasoning" and "visual" abilities. While ChatGPT was very successful at proposing productive strategies for solving the tasks on the test and expressed correct "reasoning" in most of its responses, it had difficulties correctly "seeing" graphs. We suggest that, based on its performance, caution and a critical approach are needed if one intends to use it in the role of a tutor, a model of a student, or a tool for assisting vision-impaired persons in the context of kinematics graphs.

## I. INTRODUCTION

In the last year, artificial intelligence (AI) tools, such as the Large Language Model-based chatbot ChatGPT, entered prominently into discussions around education across disciplines and educational levels. Diverse concerns have emerged regarding the impact these technologies will and, in some cases, already have, on educational practices [1,2]. Because these technologies will likely also play an important role in students' lives, scholars have argued that educational institutions should not exclude them but incorporate them into the educational process in ways that will help students learn to use them productively [1,3–5]. However, what such an integration can or should look like remains a topic of discussion and an important object of research. Furthermore, if educators want to integrate these technologies into the educational process meaningfully, it is also important for them to stay abreast of the quick technological development in the field.

In October 2023, ChatGPT acquired a new ability to process graphical input in the form of user-uploaded images. The introduction of the so-called "vision" ability opened up a range of new possibilities for its use. This new ability is particularly interesting for learning and practising physics, where multiple representations play a central and critical role. This paper aims to inform the scholarly discussion of potential educational uses of ChatGPT and similar technologies with "vision" abilities. To limit the scope of this initial investigation, the paper focuses on a particular domain of graphical representation: graphs in kinematics.

In the paper, we first provide a brief overview of how ChatGPT works and summarize studies on its performance in physics tasks. We then introduce Robert Taylors' three-role-framework to briefly discuss potential roles of ChatGPT in physics education as seen in the existing literature. To further inform such discussion in light of ChatGPT's newly acquired "vision" abilities, we then investigate ChatGPT's ability to interpret graphical input. Namely, we use the Test of understanding of graphs in kinematics (TUG-K) to probe ChatGPT's ability to interpret kinematics graphs and identify its strengths and weaknesses. Finally, we use our findings to discuss the different roles it could play in teaching and learning on graphs and suggest potential avenues for future research.

## II. BACKGROUND

### A. ChatGPT

ChatGPT is a chatbot application that allows users to interact with an AI agent through a web-based chat interface. The technology behind the chatbot is a large language model (LLM), a type of software built using machine learning algorithms [6]. An LLM has captured recurring patterns and regularities in a large training dataset consisting of natural language and other written text, such as computer code. In addition, it can also generate text based on these regularities. The text generation process is a form of statistical inference resembling an advanced "auto-complete" algorithm, where generated words are selected based on the likelihood of them appearing together in the training data and the user-provided prompt[1] [7]. Because of the enormous size of their training datasets, LLMs can typically generate convincing text in a variety of different styles and domains.

In the simplest variant of the text-generating process, the same prompt will always result in an identical output – the most likely one, according to the statistics of the training dataset and the given prompt. However, it is possible to introduce a degree of randomness into the process by increasing the value of a parameter called "temperature." In the chatbot application ChatGPT, the temperature is automatically set to a non-zero value, resulting in the chatbot giving different outputs upon being given the same prompt[2].

Research on LLMs has shown that although the process by which LLMs generate text is different from human cognition [7], their output can often resemble what we would expect from humans engaging in reasoning. This is especially true if we prompt an LLM to provide step-by-step justifications for its responses [8]. Instructing an LLM to provide a "chain of reasoning" is often referred to as *Chain-of-Thought* (CoT) prompting. Studies have shown that CoT prompting can improve the quality and correctness of responses [9,10].

We would like to caution that the term "reasoning" is an anthropomorphism, which should be used cautiously. "Reasoning" in the context of LLMs is only a human interpretation of a text output generated through a mechanism very different from actual human reasoning. Despite this, "reasoning" is often used in the research literature on LLMs [11] to refer to LLM output that presents a coherent argument resembling what would be considered a reasoning path. Similarly, "reasoning ability" refers to the ability of an LLM to produce intelligible solutions to tasks that would require reasoning when being solved by humans.

In October 2023, ChatGPT received the ability to process graphical input in the form of user-uploaded images. Because of this capability, it is sometimes also referred to as a Large Multimodal Model (LMM). This ability to "visually interpret" graphical input opens up a broad range of new possibilities and potential uses of ChatGPT. Again, the term "visual interpretation" is an anthropomorphism; the mechanisms behind machine vision employed by ChatGPT and human vision are very different.

For the sake of brevity and in line with existing literature on the topic, we will use the terms "reasoning" and "vision" as shorthands for the above-described abilities, as inferred from the analysis of ChatGPT's responses.

---

[1] The field of prompt engineering has emerged in response to the realization that having LLMs produce desired outputs is a non-trivial task that requires a focused and empirically oriented approach.

[2] Advanced users and developers can change the temperature parameter in OpenAI's application programming interface. For conversational use, non-deterministic behavior (non-zero temperature) is typically preferred.

## B. ChatGPT's performance on physics tasks

Most education research efforts on AI-based tools' performance have examined OpenAI's ChatGPT based on GPT-3.5, as well as GPT-4, which is considered the state-of-the-art of LLMs [12,13]. Researchers have already explored and reported on how different versions of ChatGPT perform on physics conceptual assessment surveys, such as the Force Concept Inventory (FCI) [14–16], how it performs on Advance Placement exams in the US [17], university physics course assessment [18], programming tasks [14], advanced data analysis [19], essay writing on the topic of history and philosophy of science [20], non-traditional conceptual physics tasks [9], as well as how it behaves in Socratic-style dialogue on introductory physics topics [21]. In this research, ChatGPT's performance was found to range from unconvincing and problematic [9,14,22], comparable to poorly performing students [15,18,22], to excellent or even expert-like [14,16,19,20]. All these explorations have focused on how ChatGPT performs on verbal, algebraic, or programming tasks. This is because ChatGPT was, until recently, only able to process written text in the form of natural or programming language and symbolic mathematical notation. In some cases, tasks involving figures were transcribed into text for ChatGPT to process them [14,16], which involved its own set of challenges.

## C. Possible roles of ChatGPT in the teaching and learning process

Existing work in education has discussed potential uses of ChatGPT and other AI-based technologies(e.g. [23,24]). We can see some common ways existing publications implicitly or explicitly frame ChatGPT's role in the education process. Here we describe three such roles, first proposed in Robert Taylor's framework of potential uses of computers in education [25]:

i. The first role can be summarized as the role of a teacher or tutor: Since the advent of ChatGPT, this was one of the most obvious potential use cases. For example, in collaboration with OpenAI, Khan Academy started a project, *Khanmigo* [26], an AI-based tutoring system based on ChatGPT. Even more recently, Google has showcased its Gemini model through a promotional video demonstrating its ability to give feedback to a student-generated solution of a physics task [27]. Researchers in PER have also started to explore how ChatGPT can be used in the role of a teacher, for example, in assessing and providing feedback on student solutions [28].

ii. The second role can be summarized as the role of a student or tutee: Existing studies often place ChatGPT in the role of a student, test its abilities, and compare them to those of human students [14,15,18,20], use it to generate synthetic response data on conceptual surveys [29], and examining its abilities to engage in Socratic dialogue with an instructor [21].

iii. The third role can be summarized as the role of a tool: This category is somewhat broader, extending beyond the educational context and the domain of physics. For example, ChatGPT can help researchers analyze research data [30] and perform other tasks associated with academic work [31]. Küchemann et al. [32] have shown that it can also support pre-service teachers in creating assessment tasks.

The above categorization, based on Taylor's three roles framework [25], is useful for three main reasons: (i) Taylor's framework is well known and established in the field of educational technology, (ii) the three roles are coherent with the existing body of research and development work in the field of AI in general and ChatGPT more particularly, and (iii) it affords meaningful discussion of educational implications of our research.

Furthermore, the three-roles framework can be used to motivate the study reported in this paper. To examine an AI-based chatbot's potential to take on any of these three roles, we need to study its performance to understand what it is capable of and what we can expect from it. If we want the chatbot to play the role of a tutor (for example, a personalized tutoring system [26,27]), we need to know how it performs on tasks that learners may ask to solve and explain. If it is to play the role of a tutee or a model of a student (for example, for teacher training [21] or for generating synthetic survey data [29]), we need to know to what extent its output resembles what could reasonably be expected from students. While the term tutee features in Taylor's original framework [25], we will mostly use the term *model of a student* instead because it better captures the use cases we present. Lastly, if we want to use it as a tool (for example, for problem-solving, such as an "object to think with" [24], as a teaching aid [28], or as an accessibility aid [33–35]), we also need to know its strengths and weaknesses, in order to delegate appropriate tasks to it.

### D. The domain of kinematics graphs: from seminal work in PER to potential uses of ChatGPT

Since the formative years of physics education research, multiple representations of physics concepts and phenomena have been central in investigating student learning. For example, graphical representations of motion, including motion diagrams and kinematics graphs, featured prominently in the University of Washington PER-group's work in the 1980s [36–38]. Research in the domain of graphs has shown that interpreting and translating between different types of graphs, including kinematics graphs, is challenging for students [39–42].

Most existing research on graphs focuses on students' skills, understanding, and challenges they experience on the topic. However, ChatGPT's new ability to interpret images expands the range of possibilities for research on the topic. What is most interesting from a physics education perspective is that ChatGPT can now be directly applied to tasks involving graphical physics representations – e.g., diagrams, sketches, and graphs. As multiple representations play a central role in the discipline of physics [43], the ability of ChatGPT to process graphical representations opens up a range of possible applications.

The topic of kinematics graphs is a good area for an initial exploration of ChatGPT's abilities for working with graphical representations. We see two main reasons for this. First, most physics students encounter kinematics graphs in their studies, regardless of the level of studies. Kinematics graphs also serve as the context in which students typically encounter graphs of physical quantities for the first time. Second, there exists a well-validated research-based survey on the topic [44,45] that can serve as a reference point for our exploration.

To have an informed discussion of ChatGPT's educational potential in the domain of kinematics graphs, we first need to get a sense of its capabilities. We have chosen to do this by investigating ChatGPT's performance on tasks that involve the interpretation of kinematics graphs. The results can serve as an initial reference point for potential comparisons of ChatGPT's with students' or experts' performance on the topic. However, more detailed insight into its strengths and weaknesses is needed to meaningfully inform if and how it can be used in the role of a tutor, a model of a student, and a problem-solving or accessibility tool. This brings us to the two research questions that we attempt to answer in this paper:

(1) How does ChatGPT perform on tasks that require interpretation of kinematics graphs?
(2) What strengths and weaknesses can be interpreted from its responses?

## III. METHODS

### A. TUG-K

To assess ChatGPT's abilities to interpret kinematics graphs, we tested its performance on the Test of Understanding Graphs in Kinematics (TUG-K)[3]. TUG-K is a multiple-choice diagnostic tool for assessing student proficiency in interpreting kinematics graphs for motion in one dimension. Initially designed by Beichner in 1994 [45], it has been updated several times, most recently in 2017 by Zavala et al. [46]. This latest version is the one we adopted for our exploration. It consists of 26 multiple-choice items designed to test seven kinematics objectives. The table in Supplemental Material [47] (reproduced from [46]) shows a list of the objectives and a description of each item that assesses them.

Together with the newest test, Zavala et al. [46] also provide a study of the performance of 471 students who took the test when enrolled in a remedial course in physics. For our purpose, we consider the student results of this study as a useful reference point for comparison to ChatGPT's performance on the test.

### B. Collection of data and coding of answers

All responses were generated using ChatGPT-4 in the "Default mode," which did not include the Python compiler plugin at the time of response generation[4]. We uploaded a "png" screenshot of each of the 26 survey items without any additional prompts. The screenshots precisely captured each TUG-K item, composed of a question, accompanying graphs, and five answer options. The whole test was submitted to ChatGPT 60 times in 1560 separate chats. Each item was always uploaded to a new conversation to avoid ChatGPT basing its answer on the text it itself generated in response to previous questions. We did not use the *regenerate* option within the same conversation precisely to avoid the chatbot deliberately changing its new response to be different from the initial one (which tends to happen when one uses the regenerate option). Still, our approach does not mean that ChatGPT generates identical responses upon repeatedly being prompted with the same question. Due to the probabilistic nature of ChatGPT's text generation and the fact that the temperature parameter is set to a non-zero value for the chatbot application, the responses are not identical upon repeated generation. For this reason, we treated them as a synthetic sample of completed surveys (see [29] for a similar approach). However, it is important to keep in mind that the responses do not reflect any specific individual's understanding of the topic. Instead, they can be seen as reflecting the different possible patterns stemming from regularities in the enormous corpus of the model's training data.

In most cases, ChatGPT provided a clear answer by explicitly stating the selected answer option (92% of responses), making the coding of its answers mostly trivial. However, in some cases, its own derived solution did not precisely match any of the provided answer options. When it did not explicitly select one of the five offered options or wrote that none were correct, we marked the answer as "not answered" (8% of responses)[5] and counted it as incorrect.

---

[3] We include only a limited number of examples of tasks from the test in this paper and its Supplemental Material. In order to facilitate the reading and interpretation of this paper, we suggest that readers obtain the whole test from the *Physport* website [44].

[4] In the rest of the paper, we refer to this model simply as ChatGPT.

[5] Almost all such responses are limited to items 4 and 16.

### C. Quantitative analysis of ChatGPT's performance on the TUG-K

To answer our first research question, the first part of the analysis looks at ChatGPT's performance on the TUG-K in terms of the selected final answer in each response. This means that we look at the final selected answer option provided by ChatGPT in its response and compare it to the test answer key, ignoring the other text accompanying it. We then compare ChatGPT's performance to those of the sample of students reported in the study by Zavala et al. [46].

We performed the analysis of the following aspects:

1) The overall performance on the test: We determine the average and median scores and the spread of the distribution of the total scores for our sample of 60 completed surveys. We quantify the spread of the distribution of scores using the interquartile range. We do this to make our results more directly comparable to the findings in [46], where the interquartile range was given instead of standard deviation due to the significantly non-normal distribution of scores.
2) Distribution of item difficulty: We look at the distribution of the number of items on the survey according to their difficulty (percentage of correct responses on the item). This allows us to compare ChatGPT's and students' performance on the test in a way that goes beyond looking just at the total score distribution.
3) Distribution of the relative frequency of selected answer options: We look at the distribution of the total number of selected answer options against the relative frequency at which they were selected. This allows us to examine how many answer options ChatGPT never or almost never selects and how many it always or almost always selects and compare this to students' responses.
4) Difficulty of survey objectives and individual items: We look at the difficulty (percentage of correct responses) of individual survey items and groups of items corresponding to survey objectives. This allows us to see which items and test objectives were the most and least difficult for ChatGPT and how this compares to students.

### D. Qualitative analysis of ChatGPT's responses on the TUG-K

To answer the second research question, the second part of the analysis looks at the content of ChatGPT's responses that precedes the final answer. We assessed the written-out responses to survey items in terms of the correctness of their (i) "reasoning" and (ii) "visual interpretation" as two separate qualities of the provided responses.

When we talk about the interpretation of graphs by humans, we tend not to focus on the distinction between their ability to interpret ("see") graphs visually and their ability to conceptually interpret the meaning of a graph in relation to mathematical and physics concepts, and real-world phenomena [42]. However, for the purpose of this paper, we need to make this distinction explicit. What we mean by "seeing" in the context of ChatGPT in this paper refers not to its ability to conceptually interpret a graph's features but instead to its ability to visually interpret the graph in terms of basic shapes, numerical values, and their spatial relations.

We thus use the term "reasoning" as a shorthand for the aspects of the answer related to argumentation based on physics and mathematical concepts and procedures. We use the term "visual interpretation" (or simply "vision") to refer to the textual descriptions of graphs provided in the responses, from which we can infer ChatGPT's ability (or inability) to "see" graphs correctly. Once again, we remind the reader that "reasoning" and "vision" are anthropomorphisms and must be used cautiously. Having acknowledged this, and for the sake of simplicity, we use the terms without quotes in the remainder of the paper.

For the qualitative analysis, we independently treat these two aspects of the responses, reasoning and visual interpretation. We do this for two main reasons. First, ChatGPT's vision ability is new and has been added as an extension to existing LLM-based abilities. Existing LLM-based abilities already include so-called reasoning abilities, which have been studied extensively [9,12,48–50]. Second, upon initial reading of the responses, we noticed that while the reasoning aspect was mostly high quality, the responses often contained faulty visual interpretation of graphs. We thus decided that a more systematic analysis of these two aspects has the potential to provide novel insights.

We analyzed one half of ChatGPT-generated responses (780 out of 1560 item responses, 30 out of 60 completed surveys) – more precisely, the first 30. Grading the quality of reasoning and visual interpretation is difficult, requiring an elaborate coding scheme. To simplify this, we decided to collapse the scoring into a binary system, where they could be correct or incorrect. The authors developed the coding scheme together through extensive dialogue after reading through the data and exploring its intricacies.

For reasoning, we marked a response as incorrect if it contained any physics, mathematics, or logic errors in the text accompanying the final answer. We also marked as incorrect those responses containing incomplete statements (e.g., “constant position implies constant velocity” is not an incorrect statement, but without explicitly stating that the constant velocity is zero in the case of constant position, the statement was considered incomplete). If the reasoning was completely absent from the response, we also marked it incorrect. Note once again that reasoning was assessed separately from vision, so a response based on a faulty visual interpretation of the graphs could still be marked as correct. These criteria are fairly conservative, as even minor errors in reasoning meant that the answer was marked as incorrect in the reasoning category.

For vision, we marked a response as incorrect if it contained an explicit mistake in the visual interpretation of the graphs. We did not mark as incorrect those responses that correctly described only some of the graphs in the task or those that did not describe all features of the graphs. So, an incorrect code was assigned to a response only in two cases: when the given descriptions contained explicit errors or if no description at all was given. This makes the coding of the correctness of visual interpretation less strict than that of reasoning.

Both authors independently coded the 780 responses. Cohen’s kappa, a measure of interrater reliability, was 0.809 (92.2% agreement) for coding reasoning and 0.967 (98.6% agreement) for coding vision. This indicates strong agreement on the coding of reasoning and almost complete agreement on the coding of vision. Most mismatches concern the coding of reasoning. This is due to errors in visual interpretation being more straightforward to notice than errors in reasoning, which were of several different types (for example, logical inconsistencies and incomplete or faulty statements). Coding for reasoning was especially cognitively demanding because it required ignoring the sometimes glaring errors in vision. All the discrepancies were resolved after a discussion between the authors, reaching a complete agreement on the coding.

## IV. FINDINGS OF THE QUANTITATIVE ANALYSIS

### A. The overall performance on the test

The average score of the sample of 60 ChatGPT-solved surveys is 10.85 points (41.7%), with the median score being 11 points. This is similar to the average (12.25, or 47%) and median (12) scores reported by Zavala et al. [46] in a sample of 471 students taking a high-school-level course in physics.

On the other hand, the interquartile range is much narrower for the chatbot. The first and third quartiles for ChatGPT data are 10 and 12 points, respectively, giving an interquartile range of only 2, as shown in Fig. 1. For students in Zavala et al.'s study [46], the first and third quartiles were 7 and 17, respectively, giving an interquartile range of 10.[6]

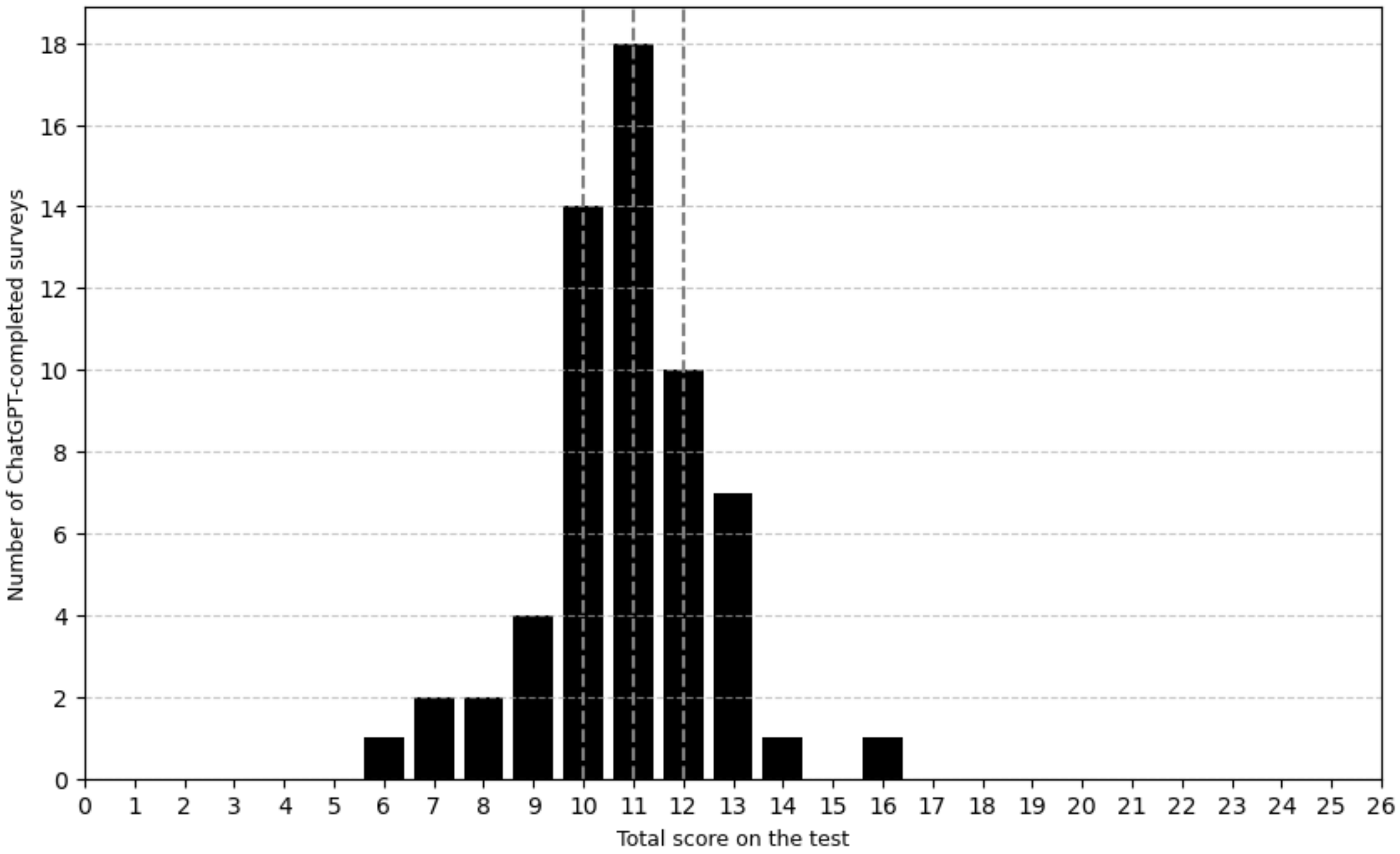

**Fig. 1:** The histogram shows the distribution of 60 scores achieved by ChatGPT on the TUG-K. The vertical dashed lines mark the quartiles for ChatGPT's distribution. Note that the distribution is much narrower than that reported in the publication by Zavala et al. [46], where Q1, Q2, and Q3 were 7, 12, and 17, respectively.

The strongest conclusion that can be drawn from the analysis so far is that ChatGPT's performance on the test as a whole is far from expert-like, in contrast to what was found in the case of the FCI [16]. Furthermore, we can see that repeated completion of the test done by the chatbot gives a much narrower distribution of scores than a group of students taking a high-school-level course in physics.

### B. Distribution of item difficulty

In Fig. 2, we can see that the number of items that ChatGPT answered correctly in less than 20% of attempts is much higher than in the student sample. The same is true for items ChatGPT got correctly in over 70% of attempts. ChatGPT's distribution of the number of items in a given difficulty range is thus different from that of students, with the extremes being much more common.

[6] The narrow spread of ChatGPT's results suggests some level of stability in the chatbot's performance, similarly to what was reported for the FCI [16].

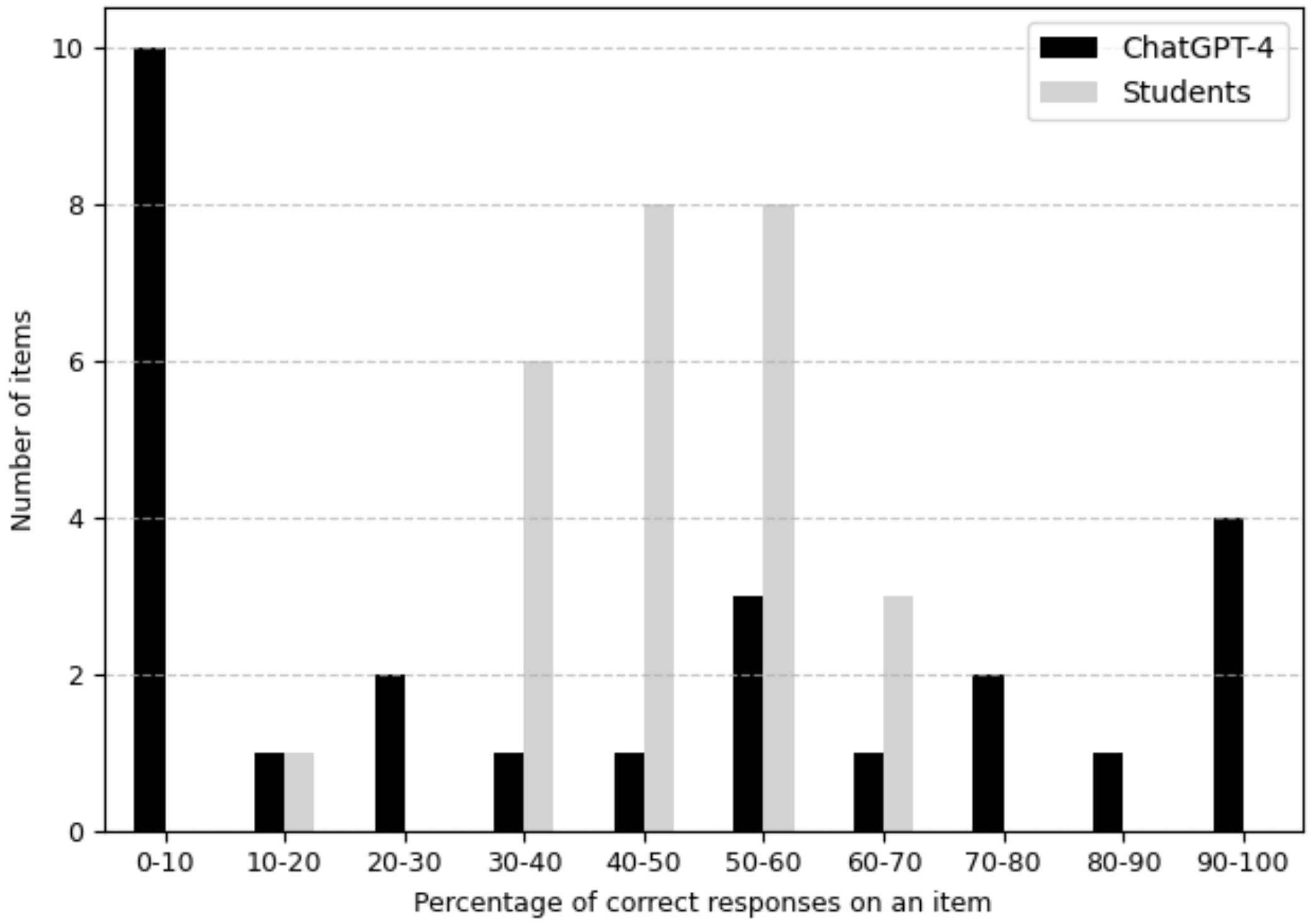

**Fig. 2:** The histogram shows the distribution of the number of items against the percentage of correct responses on an item. We can observe the overrepresentation of ChatGPT at the extremes of the distribution, meaning that, when compared to the student sample, there are more items that ChatGPT consistently (always or almost always) got right or consistently got wrong.

### C. Distribution of the relative frequency of selected answer options

A similar pattern can also be observed by looking at Fig. 3, displaying the number of selected answer options (130 answer options in total across all 26 items) vs. how often they were selected. For example, students selected 19 answer options less than 5% of the time, while for ChatGPT this number is 61. This indicates that ChatGPT is more likely to exclude certain answer options from its responses. Options selected more than 70% of the time are also overrepresented in ChatGPT's responses. Only one answer option was selected 70% of the time by students,[7] while there are 10 such answer options in ChatGPT's responses.[8] A simple interpretation of this is that ChatGPT's responses were, on average, less spread among the available answer options than those of students.

[7] Option A of item 17, which is the correct answer.
[8] This mainly happens in cases of correct answers.

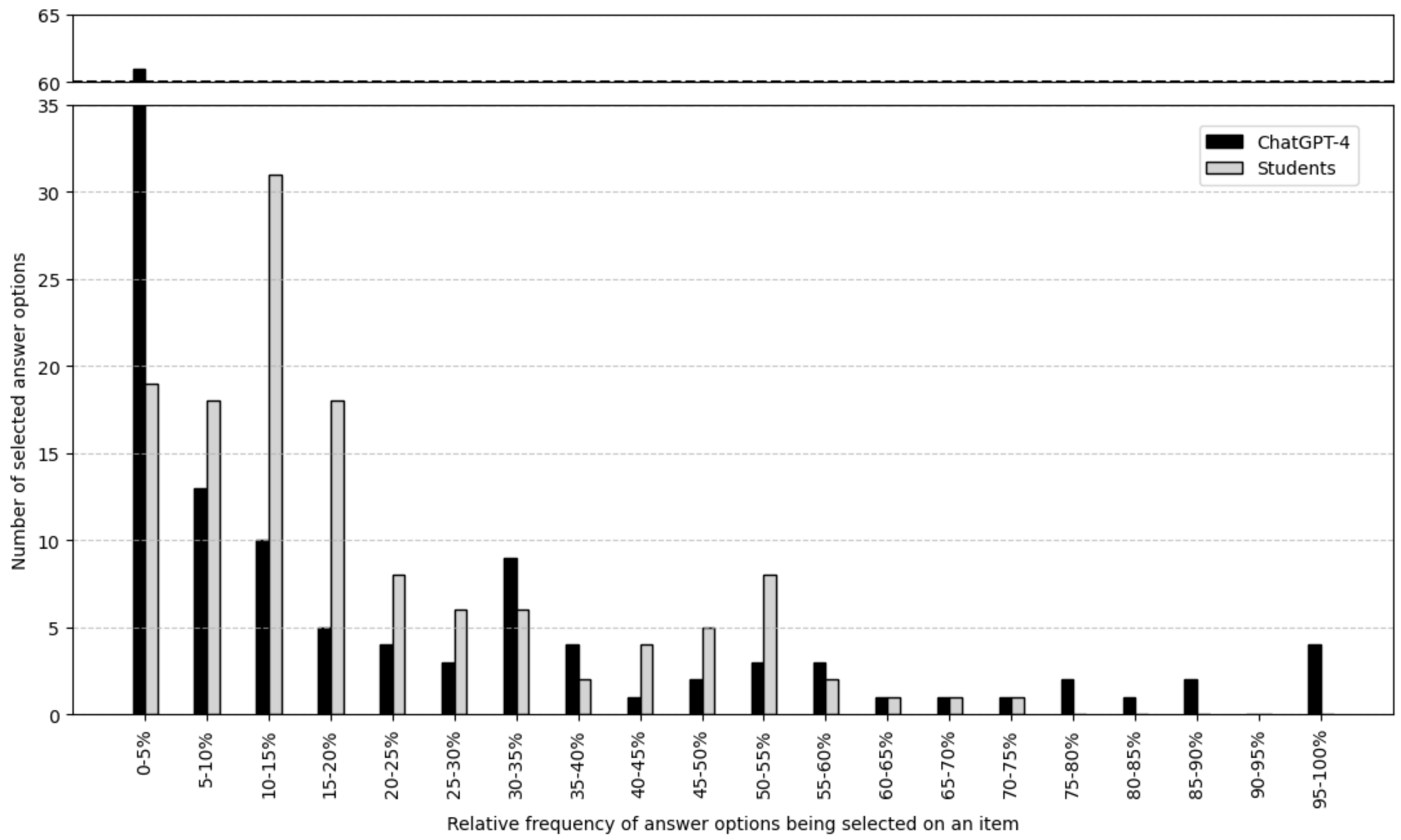

**Fig. 3:** The histogram shows the distribution of the total number of selected answer options against the relative frequency at which they were selected. We can see a clear overrepresentation of ChatGPT in the bracket of never, or almost never, selected options and an overrepresentation of ChatGPT in the brackets of options selected more than 70% of the time.

### D. Difficulty of survey objectives and individual items

Interestingly, a closer look at ChatGPT's performance on the different learning objectives assessed by TUG-K suggests that the performance of ChatGPT and students are not very different on any of the objectives, except objective 1 (*determine the velocity from the position graph*) on which ChatGPT, on average, performed 34.4 percentage points lower than students. On other objectives, the difference in performance is consistently under 20 percentage points, with ChatGPT coming on top in objectives 2 (*determine the acceleration from the velocity graph*), 3 (*determine the change of position in an interval from the velocity graph.*) and 7 (*select a graph from a textual description.*), and performing worse than students on objectives 4 (*determine the change of velocity in an interval from the acceleration graph*), 5 (*select the corresponding graph from a graph*) and 6 (*select a textual description from a graph*). However, because of the large variance of item difficulty within individual objectives (see Fig. 4), we cannot draw strong conclusions about the comparative difficulties of the different objectives.

A closer look at the performance of individual items provides a more detailed picture. In Fig. 4, we can see that ChatGPT's performance on individual items can differ markedly from that of students. Especially striking are items that ChatGPT always or nearly always got right, as well as items it almost always got wrong.

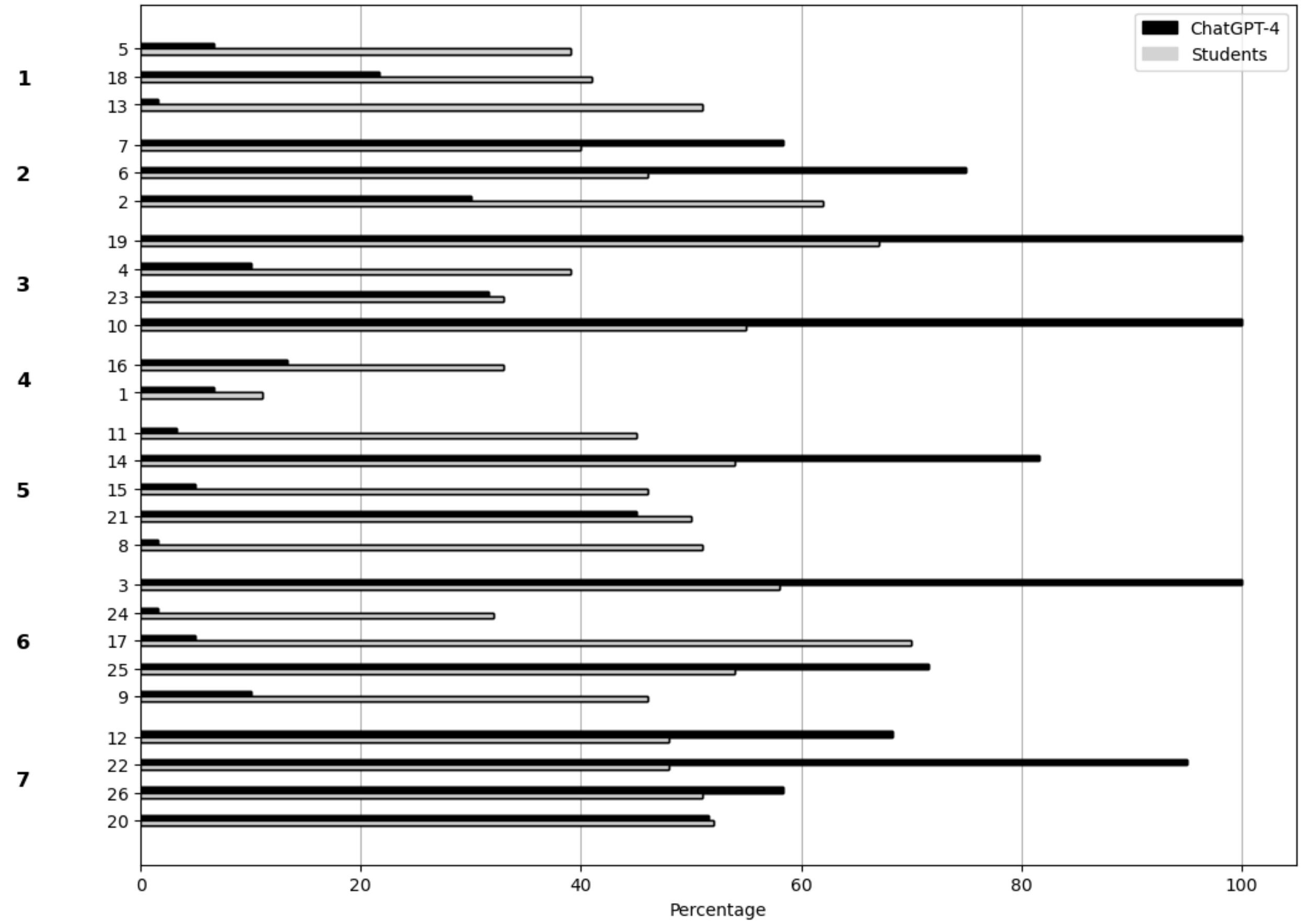

**Fig. 4:** The histogram shows the percentage of correct answers on each item on the test, grouped by the test objectives (1-7) for ChatGPT and students (as reported in [46]). For a description of items and objectives, see Appen a more detailed comparison of students' and ChatGPT's performance, see Supplemental Material [47].

In summary, the first look at ChatGPT's average and median performance on TUG-K as a whole suggests that it performs similarly to students in a remedial course in physics on the high-school level. This kind of direct comparison to student performance is tempting since it lends itself to a straightforward interpretation of ChatGPT's level of performance. Such comparisons are also commonly expressed in previous research on its performance [14,16]. However, a closer look at the spread of test results shows important differences between the ChatGPT and student samples. ChatGPT's total score distribution is much narrower than that of students. The analysis of item difficulty reveals an overrepresentation of items at both extremes of the difficulty spectrum in ChatGPT data compared to students' data. The analysis of the distribution of responses among all available answer options also reveals that ChatGPT leaves many more answer options "untouched" or "almost untouched" (5% or less). Answer options that are consistently selected (selected more than 75% of the time on a given question) are also overrepresented in the ChatGPT sample, while there are none in the student data. Moreover, important differences can be seen in the average performance on different individual items. On the other hand, strong conclusions on the test objective level cannot be made due to the large variance of the performance on items within individual objectives.

## V. FINDINGS OF THE QUALITATIVE ANALYSIS

The analysis presented above already lets us see that ChatGPT's performance on the test as a whole is far from expert-like, and also differs from what can typically be expected from students. Below, we show that this finding is further strengthened by a qualitative analysis of the content of the responses. The first noticeable difference is that even though TUG-K is a multiple-choice survey with answers marked with letters A–E, ChatGPT never provided only a letter as its answer. In contrast to what can be expected of students on the test, most of the responses start with a written strategy for solving the task. This approach to answering is also referred to in the LLM research literature as Chain-of-Thought (CoT), as mentioned in Section II.A[9]. This is also what made the type of analysis presented below possible.

### A. Correctness of reasoning and vision

Initially, we performed the Chi-square test to check the independence of the reasoning and vision variables. The test revealed no significant association between them ($\chi^2(1)=2.016$, $p=0.156$), which is in line with our assumption of their independence.

In the contingency tables collected in Table 1, we show the number of responses in each subcategory (combination of correct and incorrect for both reasoning and vision), divided into correct and incorrect answers, as well as all responses combined. We can observe in the combined contingency table that correct reasoning is present in 69.7% of all responses, while a correct visual interpretation can be found in only 30.9% of responses. This shows a clear discrepancy in the performance on these two aspects, especially given that our criteria for correct reasoning were stricter than for vision.

From the table, we can also see that not all responses with correct answers display only correct reasoning and correct vision. Perhaps most surprising is that among the responses with correct answers, 46.8% contain incorrect visual interpretations, and 16.8% contain both incorrect reasoning and visual interpretations[10]. A closer look at those responses shows that not all errors were "fatal." In some cases, it also happened that mistakes in reasoning and vision, when combined within the same response, ended up leading to the correct answer by chance. For a more detailed breakdown and examples of responses from different subcategories, see Supplemental Material [47].

When considering responses containing incorrect final answers, it is easier to understand how errors in reasoning or vision would result in a wrong answer. Here, we can see that the most prominent combination is incorrect vision and correct reasoning, making up 59.4% of responses with incorrect answers. In fact, 47% of all the analyzed responses displayed incorrect vision and correct reasoning. It is thus not surprising that we could spot this pattern even upon a quick initial reading of the responses.

In addition to the visual inspection of the contingency tables, we performed a logistic regression for all the coded data to see how good a predictor *incorrect* vision and *incorrect* reasoning are for answer *incorrectness*. The results of the logistic regression show that incorrect vision is a strong predictor of answer incorrectness ($\beta=1.887$, $OR=6.53$, $SE=0.175$, $Z=10.783$, $p<0.001$), while incorrect reasoning was a moderate predictor of answer incorrectness ($\beta=0.718$, $OR=2.05$, $SE=0.182$, $Z=3.958$, $p<0.001$). This result suggests faulty vision was the primary cause of most incorrect answers.

---

[9] We have noticed that the CoT approach to answering physics questions became prevalent in ChatGPT in summer 2023. We suspect that a system prompt utilizing CoT was implemented by OpenAI after research on the topic has shown the benefits of the approach [10].

[10] This includes 22 cases where no reasoning and/or visual interpretation were given.

**Table 1:** The contingency tables show the number of responses (out of the 780 analyzed responses) belonging to one of the four combinations of vision and reasoning correctness, first for responses with correct and incorrect answers separately, and finally for all analyzed responses together.

Responses with correct answers (327, 42%)

| Reasoning \ Vision | Incorrect | Correct | Total |
|---|---|---|---|
| Incorrect | 55 (16.8%) | 17 (5.2%) | 72 (22%) |
| Correct | 98 (30%) | 157 (48%) | 255 (78%) |
| Total | 153 (46.8%) | 174 (53.2%) | 327 |

Responses with incorrect answers (453, 58%)

| Reasoning \ Vision | Incorrect | Correct | Total |
|---|---|---|---|
| Incorrect | 117 (25.8%) | 47 (10.4%) | 164 (36.2%) |
| Correct | 269 (59.4%) | 20 (4.4%) | 289 (63.8%) |
| Total | 386 (85.2%) | 67 (14.8%) | 453 |

All responses (780, 100%)

| Reasoning \ Vision | Incorrect | Correct | Total |
|---|---|---|---|
| Incorrect | 172 (22%) | 64 (8.2%) | 236 (30.2%) |
| Correct | 367 (47%) | 177 (22.7%) | 544 (69.7%) |
| Total | 539 (69.1%) | 241 (30.9%) | 780 |

## B. Difficulties with vision

Despite seeing that ChatGPT has difficulties correctly interpreting graphs, we found that it was very reliable at reading the text in the screenshots of the test items. In other words, we have found no issues with its ability to extract the text from a "png" file correctly. From the qualitative analysis of its responses, it was clear that it successfully interpreted the task in the uploaded images. However, as indicated in the previous sub-section, we have found that it has significant issues correctly "seeing" graphs.

Our analysis of ChatGPT's responses strongly suggests that at this step, ChatGPT exhibits the most difficulty and that this is what most often sets it up for failure on the test. While we do not have a direct window into how an LMM "sees" the uploaded image, we are able to infer some features of its visual interpretation from its written responses.

Among the incorrect responses, items that are highly represented in the subcategory of incorrect vision and correct reasoning include those that require determining the surface area under given graphs (items 4, 16), comparing the surface area under given graphs (items 1, 23) determining the slope of a graph at a given point in time (items 5, 18), finding the steepest slope (item 13), and finding a matching multi-segment graph (item 15).

On the other hand, ChatGPT's reasoning seems to be much better when assessed separately from vision. The chatbot was the most successful on items explicitly asking for the right strategy to solve a given problem (items 10, 19). This further strengthens our hypothesis that it is the faulty vision, and not reasoning, that causes ChatGPT to choose the wrong answer most of the time.

One way to further test the explanation that faulty vision is causing ChatGPT to answer TUG-K items incorrectly is to transcribe graphs in the survey into text and use such transcriptions in the prompt instead of the images. This way, we bypass the need for visual interpretation and instead test only ChatGPT's ability to reason. The quick test we present here is not meant to thoroughly examine the effect of bypassing vision by transcribing the items but rather an illustrative example of the stark difference in response correctness between the two prompts.

We performed this test on item 4 (Fig. 5), on which ChatGPT, when prompted with the screenshot of the task, provided the correct overall strategy and reasoning 100% of the time but answered correctly only 10% of the time.

The response illustrates some typical vision issues we observed in the data. As the chatbot writes out its reasoning, we can notice that its description of the graph does not match the actual graph. The strategy is well delineated, but the executed reasoning is based on a faulty graph description, leading to an incorrect answer (in the case we show here, no answer was selected, which we coded as incorrect).

4. An elevator moves from the basement to the tenth floor of a building. The mass of the elevator is 1000 kg and it moves as shown in the velocity-time graph below. How far does it move during the first three seconds of motion?

A) 0.75 m
B) 1.33 m
C) 4.0 m
D) 6.0 m
E) 12.0 m

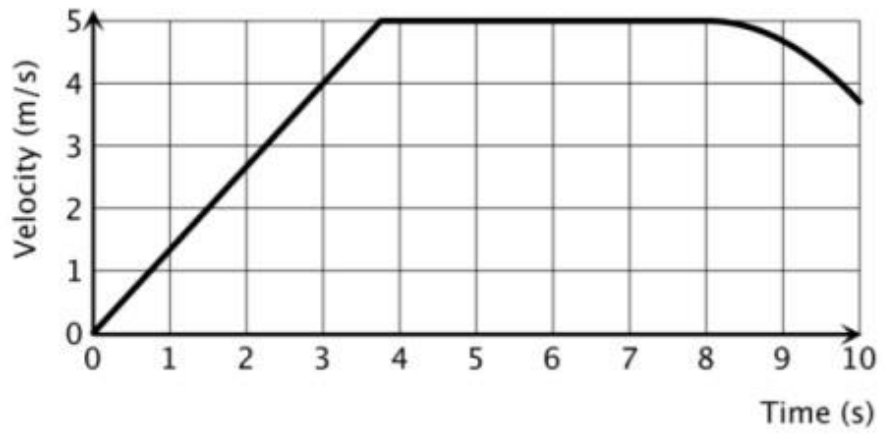

The distance an object moves is given by the area under its velocity-time graph.
For the first three seconds, the elevator's motion can be divided into two parts:
1. A triangular section from 0 to 2 seconds
2. A rectangular section from 2 to 3 seconds

1. The area of the triangle (0 to 2 seconds):
Base = 2 seconds
Height = 4 m/s (from the graph)
Area = 0.5 × base × height
Area = 0.5 × 2 s × 4 m/s = 4 m

2. The area of the rectangle (2 to 3 seconds):
Width = 1 second
Height = 4 m/s (from the graph)
Area = width × height
Area = 1 s × 4 m/s = 4 m

Adding both areas together:
Total distance = 4 m + 4 m = 8 m

So, the elevator moves 8 meters in the first three seconds. None of the given options match this calculation.

**Fig. 5:** The image shows item 4 from the TUG-K, followed by a typical ChatGPT response, where it provides a good strategy and reasoning, but fails to answer the question correctly due to faulty visual interpretation of the graph.

To test if ChatGPT could correctly solve the task if given an accurate graph description, we transcribed it and replaced the picture with the transcription. The new prompt, followed by ChatGPT's response, is shown in Fig. 6. The performance drastically improved from 10% to 100% of correctness.

An elevator moves from the basement to the tenth floor of a building. The mass of the elevator is $1000kg$ and it moves as shown in the velocity-time graph below. How far does it move during the first three seconds of motion?

**Graph description:** The graph is a velocity-time graph, with velocity on the vertical axis in units $m/s$ and time on the horizontal axis with units $s$. The graph shows a straight line from $t = 0s$, $v = 0m/s$ to $t =$ approximately $3.75s$, $v = 5m/s$, then continues horizontally at $v = 5m/s$ until $t = 8s$. Afterwards, it starts sloping downward with an increasing slope until $t = 10s$, where v= approximately $3.75m/s$.

A) $0.75m$
B) $1.33m$
C) $4.0m$
D) $6.0m$
E) $12.0m$

---

To find out how far the elevator moves in the first three seconds, we need to calculate the area under the velocity-time graph from $t = 0$ to $t = 3s$. This area corresponds to the displacement of the elevator (how far it has moved).

The graph description suggests that from $t = 0$ to $t = 3.75s$, there is a straight line increasing from $0m/s$ to $5m/s$. This indicates a constant acceleration. Since we're only considering the first 3 seconds, we will work with the part of the line that goes from 0 to 3 seconds.

Given that the line is straight and forms a right-angled triangle with the axes up to $t = 3.75s$, and $v = 5m/s$, we can calculate the area up to $t = 3s$ by finding the area of the triangle from 0 to 3 seconds and knowing the velocity at $t = 3s$.

First, we need to determine the velocity at $t = 3s$. The slope of the line, which represents acceleration (a), can be calculated using the known points $(0, 0)$ and $(3.75, 5)$.

$a =$ (change in velocity) / (change in time) $= (5 - 0)m/s/(3.75 - 0)s = 5/3.75m/s^2 = 4/3m/s^2$

Now we use this acceleration to find the velocity at $t = 3s$. We start from rest, so our initial velocity ($v_0$) is $0m/s$.

$v = v_0 + a \cdot t = 0 + (4/3) \cdot 3 = 4m/s$

So, at $t = 3s$, the velocity is $4m/s$.

Now, we calculate the displacement (which is the area of the triangle formed on the graph). The base of the triangle is the time interval ($3s$), and the height is the velocity at $t = 3s(4m/s)$.

$Area(displacement) = 0.5 \cdot base \cdot height = 0.5 \cdot 3 \cdot 4 = 6m$

So, the elevator moves 6 meters in the first three seconds. The correct answer is:

D) $6.0m$.

**Fig. 6:** The image shows the prompt created by transcribing item 4 into text, followed by a typical ChatGPT response. The response is longer, more complex, and more elaborate than what would typically be expected from a student, but it uses correct reasoning and reaches the correct conclusion. ChatGPT gave the correct answer for all the 60 repetitions of this prompt.

Although the response in Fig. 6 is more complex than what would be expected from a student (one that can see the actual graph and notice the convenient passing of the graph through the point at t=3 s and v=4 m/s), it is complete and reaches the correct conclusion. This simple test suggests that with appropriate transcriptions of graphs, ChatGPT would be able to perform better on the test. However, creating textual transcriptions for all items would be a major undertaking that goes beyond the scope of this paper. There are many challenges associated with such a project. For example, there are infinite ways of describing a graph in words. Transcription requires making decisions about the level of detail to be transcribed and necessarily involves omission or introduction of information [51]. In transcribing, one must grapple with questions, such as: "Is it important to describe how the background grid on the graph looks like?" and "To what degree is the transcription including relevant and excluding irrelevant aspects of the graph?" Further adding to the complexity of the investigation is the fact that changes in the verbal formulation of a task can have unpredictable consequences for the ChatGPT-generated response [9]. This is likely also true for pictorial input, as suggested by ChatGPT's strongly differing performance on items 5 and 6, which test a highly similar skill, but contain different graphs. It is worth noting that these two items were of comparable difficulty for students. This suggests that there are additional significant qualitative differences between ChatGPT and student interpretation of the tasks, which are potentially based on the sensitivity of the chatbot to variations in task formulation or representation, which may be trivial to students.

Another way of further testing ChatGPT's vision accuracy would be to ask it to explicitly describe provided graphs without solving the accompanying task. This approach has the potential to assess even more directly what aspects of graph visual interpretation ChatGPT has the most trouble with. Pursuing this would require the exploration of the impact of the exact textual prompt given to it together with the picture. Our previous experience with prompt engineering suggests that its performance may depend strongly on how the prompt is formulated.

### C. Difficulties with reasoning

We have seen from our analysis that the reasoning in ChatGPT's responses was predominantly of high quality. However, while less common, we could also see examples of faulty reasoning. Item 17, shown in Fig. 7, is especially interesting because it was the least difficult for students (70% correct), while being among the most difficult for ChatGPT (5% correct). Note that its visual interpretation of the graph is correct in this case. While this combination of correct visual interpretation of the graph and incorrect reasoning was uncommon in our data as a whole[11] (8%), it demonstrates that ChatGPT can be unreliable even in basic conceptual physics tasks. In fact, the reasoning in the answer is inconsistent since it expresses contradictory claims in its explanation of option D about how a velocity graph should look if there is constant, non-zero acceleration. In our experience, such behavior is quite atypical of students and would be hard to understand if we assume that the answer was generated by a human. The responses generated by the chatbot often have this uncanny quality [21].

Referring back to the possibility of improving ChatGPT's performance by transcribing graphs into language, this example is a reminder that even when graphs are correctly visually interpreted by ChatGPT, this does not guarantee correct reasoning or a correct answer. This corroborates our previous findings on the unreliability of the performance of ChatGPT on conceptual physics tasks [9,21] and carries significance for its potential applications in education.

[11] Though not uncommon on this particular item.

17 . The graph to the right shows the velocity of an object moving in a straight line. Which sentence is the best interpretation?

A) The object is moving with a uniformly increasing position.
B) The object´s position is constant.
C) The object is moving with a uniformly increasing acceleration.
D) The object is moving with a constant, non-zero acceleration.
E) The object is moving with a uniformly increasing velocity.

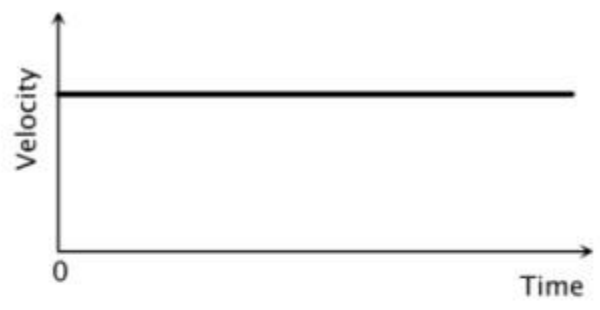

The graph shows a constant, horizontal line for velocity with respect to time. This means that the velocity of the object is not changing over time. Given the options:

A) Incorrect. A uniformly increasing position would show a velocity graph with a positive, constant slope, not a horizontal line.

B) Incorrect. A constant position means the object is not moving, so its velocity would be zero. However, this graph shows a constant, non-zero velocity.

C) Incorrect. A uniformly increasing acceleration would show a velocity graph with a constantly increasing slope, not a horizontal line.

D) Correct. A constant, non-zero acceleration would change the velocity over time. However, a constant velocity, as shown by a horizontal line, indicates the object is moving with a constant, non-zero acceleration.

E) Incorrect. A uniformly increasing velocity would show a velocity graph with a positive, increasing slope, not a horizontal line.

So, the best interpretation is: D) The object is moving with a constant, non-zero acceleration.

**Fig. 7:** The figure shows item 17 from the TUG-K, followed by a typical response from ChatGPT. On this item, ChatGPT always correctly visually interpreted the graph but consistently presented incorrect reasoning.

## VI. DISCUSSION

### A. Summary of findings

Our two research questions are addressed by the quantitative and qualitative parts of our analysis, respectively.

(1) How does ChatGPT perform on tasks that require interpretation of kinematics graphs?

To answer this question, we investigated ChatGPT's performance on the TUG-K. Our quantitative analysis found that ChatGPT's performance is far from expert-like. Its average and median performance is comparable to a sample of students taking a physics course at the high school level [46]. However, the distribution of responses is different. First, ChatGPT's performance distribution was less spread out, having an interquartile range of only 2, compared to 10 in the student data. Second, ChatGPT more consistently answered correctly or incorrectly on specific test items, as seen in the much larger number of items with the average score of less than 10% or more than 75% in the ChatGPT data compared to the student data. Third, ChatGPT left many more answer options wholly or nearly "untouched." This includes both distractors and correct answers.

(2) What strengths and weaknesses can be interpreted from its responses?

Analyzing a sample of 780 responses for the quality of reasoning and visual interpretation, we saw that ChatGPT provides correct reasoning in 69.7% of responses. At the same time, we can see that its visual interpretation of the graphs is without errors in only 30.9% of responses. Consequently, we suggest that vision is a major reason ChatGPT answers the questions incorrectly. This is supported by a logistic regression of the data on vision incorrectness and the incorrectness of the responses. Furthermore, a quick test, where we interpreted and transcribed one graphical survey item into text, suggests that ChatGPT handles textual descriptions better than image input, further supporting our hypothesis that incorrect vision, and not incorrect reasoning, is a major reason for its incorrect answers.

While its reasoning capabilities appear to be better and more reliable than the visual interpretation of graphs, we have still found that ChatGPT sometimes provides incorrect reasoning. This is in agreement with our earlier findings on its unreliability on conceptual physics tasks [9]. Its responses can also contain unusual mistakes and inconsistencies that can appear uncanny and hard to understand for educators used to working with human students.

In summary, ChatGPT's main strength is the ability to provide correct and appropriate strategies for solving the tasks on TUG-K. In contrast, its main weakness is its inability to visually interpret graphs correctly.

## B. ChatGPT's three potential roles

Here, we return to the three roles framework [25] proposed in Section II and use it to discuss how the findings reported in the paper can inform the use of ChatGPT. We then address ideas for future research and discuss the study's limitations.

### *1. ChatGPT as a tutor*

We have seen that ChatGPT does not perform on the TUG-K at the level that, according to our belief, should be expected of a well-prepared physics teacher. Its average performance on the test already suggests that it would not be a good tutor on the topic. However, certain aspects of its responses permit the possibility of it being useful to help students learn to deal with graph-related tasks. Using a Chain-of-Thought approach, we saw that it mostly provided appropriate and well-formulated strategies for approaching different tasks. For students, such big-picture strategies could be framed as potential approaches to solving problems that need further exploration and evaluation by students with the support of a human instructor.

On the other hand, its lousy performance in the visual interpretation of graphs presents a major limitation here. The faulty descriptions of graphs can potentially confuse learners, and we would therefore advise that the image recognition function in its current form is not appropriate for tutor-like applications of ChatGPT in the domain graphs.

Lastly, given that 30.2% of responses had faulty or completely missing reasoning, we would suggest that ChatGPT should not be relied on in physics reasoning tasks either. This supports our earlier findings that ChatGPT still experiences issues with conceptual reasoning on introductory physics topics [9]. All in all, the unreliability of its performance in reasoning and vision renders it untrustworthy enough that we would advise against its use as a tutor when not supervised by an experienced human teacher.

### 2. *ChatGPT as a model of a student*

Given the similarity of ChatGPT's and students' average performance on the TUG-K, one could conclude that ChatGPT can serve as a good model for students at the high school level. However, as we have shown both through a more detailed quantitative and qualitative analysis, it differs from students in several ways. Looking at the test as a whole, it should not be used to simulate the performance of hypothetical samples of students (generating synthetic survey-response data [29]), for example, to investigate the difficulty and other characteristics of surveys and tests similar to TUG-K. The same is true on the item level. We have seen that ChatGPT was more consistently right and wrong on specific items and more frequently and consistently "avoided" more of the distractors (and sometimes even correct answers). However, as [29] has shown, special prompting approaches can be a potential way to have ChatGPT generate responses that more closely resemble common student difficulties. We have not applied any specific textual prompts in this study besides the text in the image inputs (screenshots of the tasks).

Qualitatively, ChatGPT used a Chain-of-Thought approach to answering the survey items. We do not typically expect students to write their reasoning on multiple-choice assessments. However, the CoT approach provides us with insight into how the chatbot reaches its conclusions. This was also found to be different from students. As we have seen, ChatGPT proposed a clear and well-articulated strategy and reasoned correctly in most of the analyzed responses but failed on many tasks due to difficulties in correctly visually interpreting the graphs in the tasks. In our experience, the challenge with human students is often reversed. Most can see the graphs well, but those who fail at such tasks typically do not know, or have a wrong understanding of, what they should do to arrive at the correct answer. It seems unlikely that students who could express the strategies and explanations as clearly as ChatGPT would perform as poorly on the test as it did. To further explore this, it would be valuable to explore student reasoning in more detail by asking them to explain their reasoning in a CoT fashion, for example, through think-aloud protocols.

Thus, we suggest that ChatGPT is mostly unsuitable for serving as a model of individual students experiencing typical learning difficulties, on which teachers could train their pedagogical skills. Furthermore, as we have seen when ChatGPT exhibits difficulties in reasoning, they often feel "uncanny" and not qualitatively similar to students' difficulties that we have experienced in our work as high school and university teachers.

### 3. *ChatGPT as a tool*

The above-discussed limitations present serious drawbacks to ChatGPT's use as a problem-solving tool in physics. While the output of ChatGPT, or any other AI-based tool, should always be evaluated carefully, this seems especially true if used for interpreting graphical input. Recognizing its failure to correctly visually interpret image input is especially important when considering it as a potential assistive tool for vision-impaired persons. This is true both for learning and professional contexts. We suggest that alternative approaches to conveying graphs to vision-impaired persons should be considered until the reliability of ChatGPT or similar technologies on visual interpretation is significantly improved.

Given its unreliability also in reasoning, especially on conceptual questions [9], ChatGPT still appears not to be a good tool for outsourcing physics conceptual reasoning. However, for more calculation-heavy tasks, this seems to be changing with the advent of the *Advanced Data Analysis* mode (previously known as the *Code Interpreter* plugin) [19]. Given that in our study, ChatGPT mostly gave good suggestions of problem-solving strategies, there is undoubtedly real potential for its productive use, as long as the users, be they students or teachers, know its limitations and delegate tasks to it in ways that leverage its strengths and bypass its weaknesses. Doing this requires a critical approach to interpreting AI-generated output. Activities aimed at improving students' critical thinking and reading skills in relation to AI and physics content (e.g., [52]) will likely play an increasingly important role in preparing students for the future.

### C. Future work

This paper presents one of the first studies examining the performance of an LMM-based chatbot, ChatGPT, on physics tasks that require the interpretation of graphical input. While the TUG-K survey is focused on kinematics graphs, many other areas of physics involve some form of pictorial or diagrammatic representation, such as force diagrams, Feynman diagrams, circuit diagrams, etc. The performance of ChatGPT on surveys, such as the FCI, has previously been studied [14–16], but not using the recently introduced image input functionality. We believe that exploring LMMs' performance on visually-heavy tasks is an exciting and upcoming area of research, with significance for understanding and fruitfully applying these new technologies in physics education and physics education research. More generally, different prompt engineering techniques, applied to both text and images, may be one way to improve the performance [9,53] or make it resemble student data more closely [29]. In our work, we did not test if using more advanced prompting techniques in combination with image input might generate more student-like responses, so this remains a possible area for future research.

This paper only provides a momentary snapshot of the capabilities of the first widely available LMM at a point in time very close to its public release. In the future, continuous assessment of these models will be necessary if the physics education community is to stay abreast of the rapid technological development in the field of generative AI. We have also made our data available in an open online repository as a resource for supporting potential future studies [54].

### D. Limitations

As mentioned above, it is important to keep in mind that AI technology is developing extremely fast. This means that this paper presents only a snapshot of ChatGPT's abilities as they were in October 2023. Given that OpenAI is regularly updating its models, it is unlikely that the performance will remain the same for a long time. The sensitivity of models like ChatGPT to seemingly minor prompt changes means this study's findings are not automatically generalizable to tasks beyond the TUG-K. The study did not systematically explore what features of graph presentation are most challenging to interpret, so it cannot provide guidance on what image prompts are likely to produce the best outcomes. The possibility of generalizing our findings is also limited regarding other representations, such as force and electric circuit diagrams. Studies of those contexts are needed to see if similar patterns are present.

## VII. CONCLUSION

The study presented in this paper found that ChatGPT's ability to interpret kinematics graphs, as measured by the TUG-K, is far from expert-like. It is, on average, roughly comparable to the performance of students taking a high-school physics level course, but with important differences in other aspects, such as the spread of the distribution of total scores and the distribution of item difficulty. Our qualitative analysis of ChatGPT's responses reveals that its main strength is the ability to provide correct strategies for solving the problems on the test. However, despite this, most of its answers were incorrect. The main culprit for this is its inability to correctly visually interpret graphs. While the findings presented in the paper are largely limited to the TUG-K survey, they remind us to be cautious when considering using ChatGPT in other tasks involving image processing. We hope that the paper can offer a possible way of approaching the study of visual tasks also in other domains of physics.

# APPENDIX A

Description of objectives and items of the modified version of TUG-K from Zavala *et al.* (Modifying the test of understanding graphs in kinematics, Phys. Rev. Phys. Educ. Res. 13, 020111 (2017)).

| Objectives | Items | Description |
|---|---|---|
| 1 – Determine the velocity from the position graph. | 5 | Determine the positive value of velocity in a time from the position graph. |
| | 18 | Determine the negative value of velocity in a time from the position graph. |
| | 13 | Determine the interval with most negative velocity from the position graph. |
| 2 – Determine the acceleration from the velocity graph. | 7 | Determine the positive value of acceleration in a time from the velocity graph. |
| | 6 | Determine the negative value of acceleration in a time from the velocity graph. |
| | 2 | Determine the interval with most negative acceleration from the velocity graph. |
| 3 – Determine the change of position in an interval from the velocity graph. | 19 | Establish the procedure to determine the change of position in an interval from the velocity graph. |
| | 4 | Determine the change of position in an interval from the velocity graph. |
| | 23 | Determine the greatest change in position in an interval from the velocity graph. |
| | 10 | Establish the procedure to determine the change of velocity in an interval from the acceleration graph. |
| 4 – Determine the change of velocity in an interval from the acceleration graph. | 16 | Determine the change of velocity in an interval from the acceleration graph. |
| | 1 | Determine the greatest change in velocity in an interval from the acceleration graph. |
| 5 – Select the corresponding graph from a graph. | 11 | Select the velocity graph from the position graph. |
| | 14 | Select the acceleration graph from the velocity graph. |
| | 15 | Select the velocity graph from the acceleration graph. |
| | 21 | Select the position graph from the velocity graph. |
| | 8 | From the position graph determine that the movement of an object is as follows: it does not move, moves backwards and then stops. |
| 6 – Select a textual description from a graph. | 3 | From the position graph determine that the object moves at constant velocity. |
| | 24 | From the velocity graph determine that the object moves at constant acceleration. |
| | 17 | From the velocity graph determine that the object increases its position uniformly. |
| | 25 | From the acceleration graph determine that the object increases its velocity uniformly. |
| | 9 | Identify the position graph that corresponds to a positive and constant acceleration. |
| 7 – Select a graph from a textual description. | 12 | Identify the position and velocity graphs that correspond to a constant velocity. |
| | 22 | Identify the velocity and acceleration graphs that correspond to a constant nonzero acceleration. |
| | 26 | Identify the velocity and acceleration graphs that correspond to a velocity that increases uniformly. |
| | 20 | Identify the acceleration graph that corresponds to an acceleration that increases uniformly. |

# APPENDIX B

Performance on the TUG-K of ChatGPT and students, reported as percentage. The correct answers are in bold. Students' performance is reproduced from Zavala *et al.* (Modifying the test of understanding graphs in kinematics, Phys. Rev. Phys. Educ. Res. 13, 020111 (2017)).

| Objective | Item | | A | B | C | D | E | N |
|---|---|---|---|---|---|---|---|---|
| 1 | 5 | GPT | 3,3 | 0 | **6,6** | 58,3 | 30 | 1,6 |
| | | STU | 5 | 1 | **39** | 42 | 12 | 1 |
| | 18 | GPT | **21,6** | 51,6 | 8,3 | 11,6 | 3,3 | 3,3 |
| | | STU | **41** | 13 | 11 | 23 | 11 | 1 |
| | 13 | GPT | **1,6** | 0 | 61,6 | 36,6 | 0 | 0 |
| | | STU | **51** | 14 | 2 | 31 | 1 | 0 |
| 2 | 7 | GPT | **58,3** | 10 | 31,6 | 0 | 0 | 0 |
| | | STU | **40** | 19 | 24 | 13 | 3 | 1 |
| | 6 | GPT | 11,6 | **75** | 10 | 0 | 0 | 3,3 |
| | | STU | 18 | **46** | 11 | 12 | 11 | 2 |
| | 2 | GPT | 36,6 | 31,6 | 1,6 | 0 | **30** | 0 |
| | | STU | 2 | 14 | 20 | 1 | **62** | 0 |
| 3 | 19 | GPT | 0 | **100** | 0 | 0 | 0 | 0 |
| | | STU | 2 | **67** | 13 | 15 | 3 | 0 |
| | 4 | GPT | 0 | 0 | 6,6 | **10** | 0 | 83,3 |
| | | STU | 4 | 11 | 15 | **39** | 28 | 2 |
| | 23 | GPT | 15 | **31,6** | 0 | 30 | 8,3 | 15 |
| | | STU | 3 | **33** | 18 | 19 | 26 | 1 |
| | 10 | GPT | **100** | 0 | 0 | 0 | 0 | 0 |
| | | STU | **55** | 28 | 4 | 6 | 7 | 1 |
| 4 | 16 | GPT | 6,6 | 0 | 3,3 | **13,3** | 0 | 76,6 |
| | | STU | 12 | 27 | 12 | **33** | 15 | 0 |
| | 1 | GPT | **6,6** | 20 | 1,6 | 6,6 | 36,6 | 28,3 |
| | | STU | **11** | 10 | 1 | 54 | 23 | 1 |

| Objective | Item | | A | B | C | D | E | N |
|---|---|---|---|---|---|---|---|---|
| 5 | 11 | GPT | 33,3 | 6,6 | 15 | **3,3** | 41,6 | 0 |
| | | STU | 12 | 27 | 12 | **33** | 15 | 0 |
| | 14 | GPT | 0 | **81,6** | 18,3 | 0 | 0 | 0 |
| | | STU | 10 | **54** | 7 | 16 | 12 | 1 |
| | 15 | GPT | **5** | 28,3 | 38,3 | 20 | 8,3 | 0 |
| | | STU | **46** | 6 | 17 | 11 | 21 | 0 |
| | 21 | GPT | 45 | **45** | 0 | 8,3 | 1,6 | 0 |
| | | STU | 8 | **50** | 6 | 19 | 17 | 0 |
| | 8 | GPT | 86,6 | 0 | 0 | **1,6** | 10 | 1,6 |
| | | STU | 10 | 14 | 14 | **51** | 11 | 0 |
| 6 | 3 | GPT | 0 | 0 | 0 | **100** | 0 | 0 |
| | | STU | 9 | 2 | 29 | **58** | 3 | 0 |
| | 24 | GPT | **1,6** | 76,6 | 21,6 | 0 | 0 | 0 |
| | | STU | **32** | 42 | 15 | 8 | 1 | 3 |
| | 17 | GPT | **5** | 53,3 | 0 | 16,6 | 18,3 | 6,6 |
| | | STU | **70** | 10 | 3 | 14 | 4 | 0 |
| | 25 | GPT | 0 | 1,6 | **71,6** | 26,6 | 0 | 0 |
| | | STU | 5 | 8 | **54** | 23 | 6 | 4 |
| | 9 | GPT | 88,3 | 0 | 0 | 1,6 | **10** | 0 |
| | | STU | 6 | 32 | 10 | 6 | **46** | 0 |
| 7 | 12 | GPT | 0 | **68,3** | 0 | 0 | 31,6 | 0 |
| | | STU | 7 | **48** | 18 | 15 | 11 | 0 |
| | 22 | GPT | 3,3 | 0 | **95** | 0 | 0 | 1,6 |
| | | STU | 12 | 20 | **48** | 11 | 8 | 0 |
| | 26 | GPT | 0 | **58,3** | 1,6 | 30 | 10 | 0 |
| | | STU | 6 | **51** | 3 | 19 | 15 | 5 |
| | 20 | GPT | 28,3 | 13,3 | **51,6** | 5 | 1,6 | 0 |
| | | STU | 2 | 15 | **52** | 21 | 9 | 0 |

# APPENDIX C

## CORRECT ANSWERS

In 327 cases out of 780 (42%) of the coded responses, ChatGPT gave the correct answer. This also roughly matches the overall average performance of ChatGPT on the test as a whole (41.7%). However, our analysis shows that even for responses containing correct answers ChatGPT's displayed visual interpretation and reasoning capabilities were far from perfect. Among the responses containing correct answers, 46.8% displayed faulty vision and 22% displayed faulty reasoning. Still, the largest subcategory, containing 48% of correct answers, was that displaying correct reasoning and correct vision. Below, we provide illustrative examples of responses for each of the four subcategories of correct answers.

**Correct vision, correct reasoning**

This is the most populated subcategory of correct responses. While we initially expected that most correct answers would fall into this subcategory, it contains only 48% of the responses with correct answers.

22. Consider the following graphs, noting the different axis:

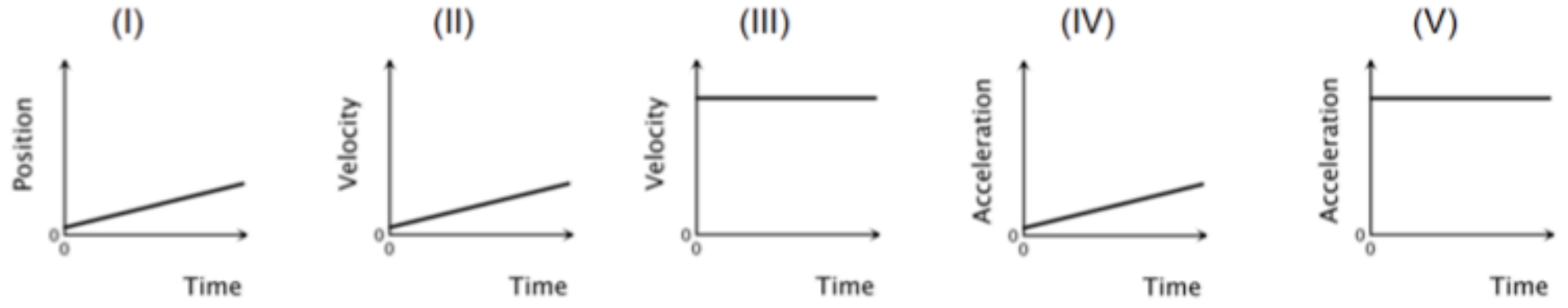

Which of these represent(s) the object´s motion with a constant, non-zero acceleration?

A) I, II and IV
B) V only
C) II and V
D) IV only
E) III and V

Let's break down each graph:

(I) Position vs. Time: This graph shows a linear increase in position over time, which means there's a constant velocity. However, for constant acceleration, the position-time graph should be a curve (quadratic). This does not represent constant, non-zero acceleration.
(II) Velocity vs. Time: This graph shows a linear increase in velocity over time, which indicates a constant acceleration. This represents constant, non-zero acceleration.
(III) Velocity vs. Time: The velocity remains constant over time. This means there's zero acceleration, not a constant non-zero acceleration.
(IV) Acceleration vs. Time: This graph shows an increasing acceleration over time. This means the acceleration is not constant.
(V) Acceleration vs. Time: This graph indicates a constant acceleration over time. It remains at a value above zero, which means it's a constant, non-zero acceleration.

From the analysis:
- (II) and (V) represent motion with constant, non-zero acceleration.
The correct answer is C) II and V.

Figure 8: The example shows a case of what could be called exemplary performance. We can infer from the response that ChatGPT correctly visually interpreted the graphs (good vision), and reasoned well about them (good reasoning).

**Incorrect vision, correct reasoning**

In this subcategory, we mostly see responses where the errors in vision were such that the answer still turned out correct. This often happened in a way where ChatGPT chose a response that was the most similar or numerically closest to the solution it arrived at through correct reasoning.

7. The graph shows the velocity in function of time of an object that is moving in a straight line. At $t$ = 65 s the instantaneous acceleration of the object was most nearly:

A) 1.0 m/s$^2$
B) 2.0 m/s$^2$
C) 0.46 m/s$^2$
D) 30 m/s$^2$
E) 34 m/s$^2$

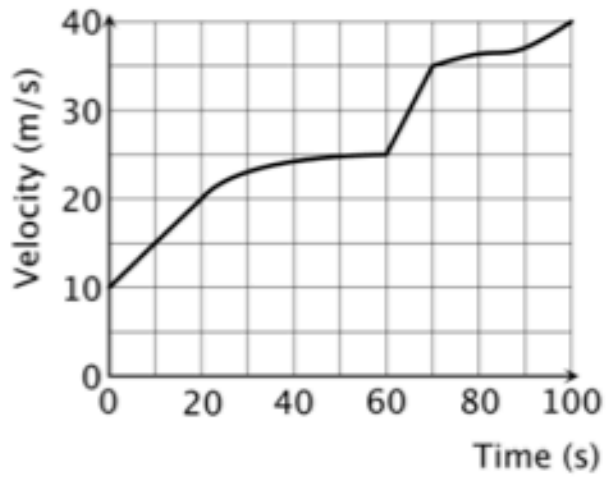

The instantaneous acceleration of an object at a specific time can be found by determining the slope of the velocity-time graph at that point. The steeper the slope, the greater the acceleration.

To determine the acceleration at $t = 65$ s, we'll want to find the slope of the tangent line to the graph at that time. Given the nature of the image, we have to make an estimation.

Around $t = 65$ s, the velocity seems to be increasing at a relatively constant rate over a short interval. Let's estimate the change in velocity over a small interval near $t = 65$ s.

Using the grid as a guide, it appears that **between approximately $t = 60$ s and $t = 70$ s, the velocity increases from roughly 30 m/s to roughly 38 m/s.** This is an increase of 8 m/s over 10 seconds.

Using the formula for acceleration:

$$a = \frac{\Delta v}{\Delta t}$$

Where:
- $\Delta v$ is the change in velocity, and
- $\Delta t$ is the change in time.

$$a = \frac{8m/s}{10s} = 0.8m/s^2$$

From the given options, the closest to 0.8 m/s$^2$ is:
A) 1.0 m/s$^2$.

Figure 9: The example shows a case of correct reasoning, but faulty visual interpretation of the graph. The part of the response from which we inferred faulty vision is bolded. On the graph, the values of velocity at $t = 60s$ and $t = 70s$ are $25m/s$ and $35m/s$, respectively, and not $30m/s$ and $38m/s$. The error in vision is such that the calculation is close to the correct answer.

**Incorrect vision, incorrect reasoning**

We were somewhat surprised to see that 16.8% of correct answers fell into this subcategory. This is partially a consequence of our coding, where we coded responses where no visual interpretation or reasoning could be inferred from them as incorrect. Another group of responses that ended up in this subcategory were those where mistakes in vision and reasoning were identified, but were not crucial for arriving at the answer. In some cases, a correct answer was a result of inconsistent and difficult-to-understand conclusions, made on the basis of faulty vision and reasoning, resulting in correct answers "by coincidence". The last reason for a response ending up in this subcategory is when two mistakes "cancel each other out", like in the example shown below.

12. Consider the following graphs, noting the different axes:

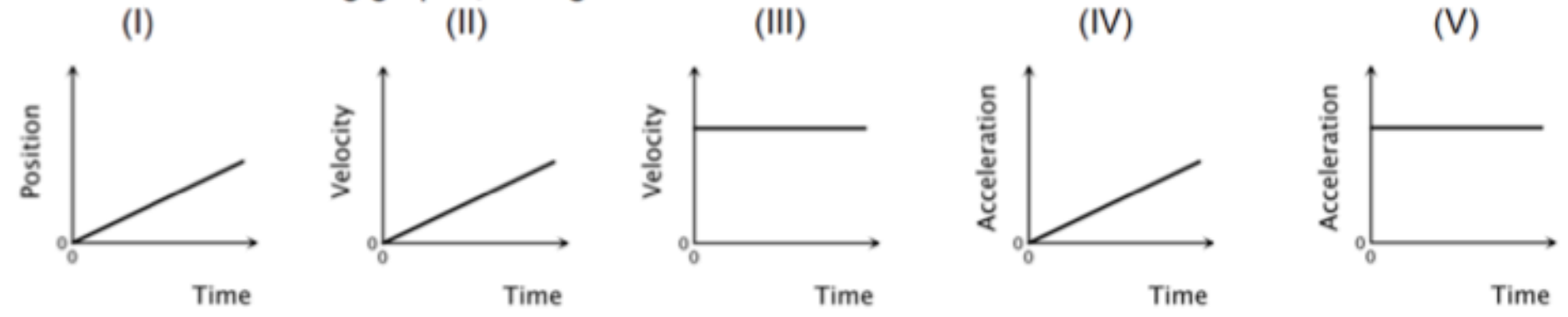

Which of these represent(s) motion at constant velocity?

A) I, II and IV.
B) I and III.
C) III only.
D) III and V.
E) I, III and V.

For motion at constant velocity:
- The position vs. time graph would be a straight line.
- The velocity vs. time graph would be a horizontal line (indicating no change in velocity).
- The acceleration vs. time graph would be a horizontal line at the 0 mark (indicating no acceleration).

Based on the graphs:
- (I) is a straight line in the position vs. time graph, indicating constant velocity.
- (II) has an increasing velocity, so it's not constant velocity.
- (III) has constant velocity as seen in the velocity vs. time graph.
- (IV) shows a positive acceleration, meaning velocity is not constant.
- (V) **has zero acceleration throughout, but acceleration being zero doesn't necessarily mean the motion was at constant velocity**, just that it wasn't changing during that time. We'd need more information to determine if the velocity was initially constant.

From the choices:
B) I and III. are the graphs that represent motion at constant velocity.

Figure 10: The example shows a case of faulty vision combined with faulty reasoning (marked in bold), ending up in the choice of the correct answer.

**Correct vision, incorrect reasoning**

This was the least populated of all the subcategories, containing only 5.2% of responses with correct answers. It contains responses from which we can infer correct vision, but contain faulty reasoning, or where the reasoning is missing altogether (3 cases).

25. The graph to the right represents the acceleration as a function of time for a moving object. Which sentence is the best interpretation?

A) The object is moving with a uniformly increasing acceleration.
B) The object does not move.
C) The object is moving with a uniformly increasing velocity.
D) The object is moving at a constant velocity.
E) The object is moving with a uniformly increasing position.

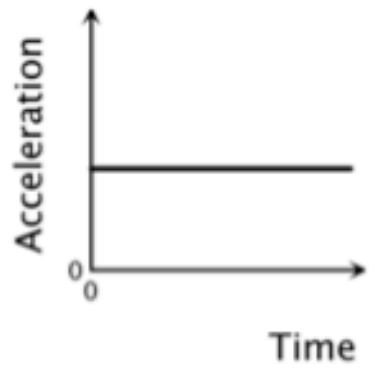

The graph shows that the acceleration of the object is constant over time (since the acceleration remains the same value on the y-axis as time progresses on the x-axis).

Given the options:
A) Incorrect, as the acceleration is not changing over time.
B) Incorrect, just because acceleration is constant doesn't mean the object is not moving.
C) Correct, a constant acceleration means the velocity of the object is increasing uniformly over time.
D) Incorrect, constant velocity would mean zero acceleration.
E) **Incorrect, uniformly increasing position would imply an increasing slope for velocity and thus a constant acceleration**, but it is more accurately described in option C.

So, the best interpretation is:
C) The object is moving with a uniformly increasing velocity.

Figure 11: The example shows a response containing a correct visual interpretation of the graph, but which also contains wrong reasoning, marked in bold.

## INCORRECT ANSWERS

In 453 cases out of 780 coded responses (58%), ChatGPT gave an incorrect answer or did not choose any of the available answer options (both coded as incorrect). Of these responses, only 14.8% contained correct visual interpretation. The rest were either faulty (vast majority), or no visual interpretation could be inferred from the response. Interestingly, 63.8% of the responses coded as incorrect displayed good reasoning. Not surprisingly, the largest subcategory among responses with incorrect answers was that of incorrect vision and correct reasoning. Below, we provide illustrative examples of responses for each of the four subcategories of responses coded as incorrect.

**Incorrect vision, correct reasoning**

This was the biggest subcategory among responses, providing an incorrect answer (59.4%), and among all analysed responses (47%). It reflects the overall pattern of predominantly correct reasoning and predominantly incorrect visual interpretation, suggesting thus that wrong answers mainly follow from mistakes in vision.

4. An elevator moves from the basement to the tenth floor of a building. The mass of the elevator is 1000 kg and it moves as shown in the velocity-time graph below. How far does it move during the first three seconds of motion?

A) 0.75 m
B) 1.33 m
C) 4.0 m
D) 6.0 m
E) 12.0 m

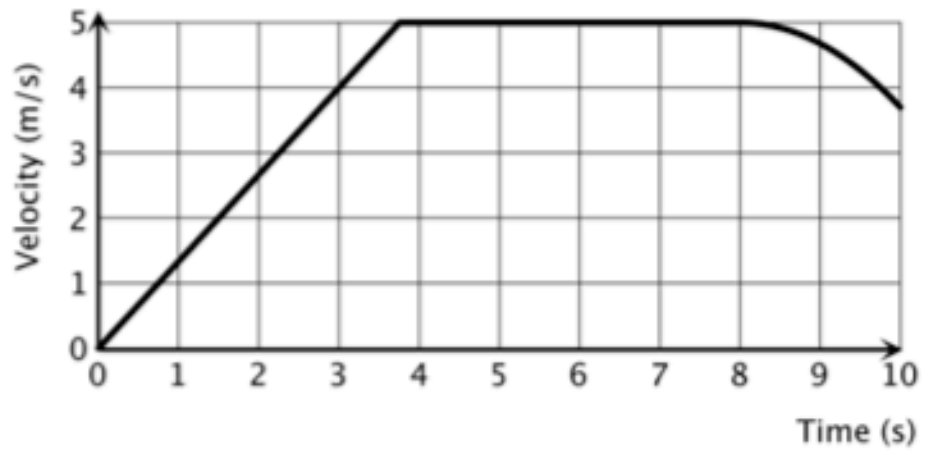

The distance an object moves is given by the area under its velocity-time graph.
For the first three seconds, the elevator's motion can be divided into two parts:
1. A triangular section from 0 to 2 seconds
**2. A rectangular section from 2 to 3 seconds**

1. The area of the triangle (0 to 2 seconds):
Base = 2 seconds
**Height = 4 m/s (from the graph)**
Area = 0.5 × base × height
Area = 0.5 × 2 s × 4 m/s = 4 m

2. The area of the rectangle (2 to 3 seconds):
Width = 1 second
**Height = 4 m/s (from the graph)**
Area = width × height
Area = 1 s × 4 m/s = 4 m

Adding both areas together:
Total distance = 4 m + 4 m = 8 m

So, the elevator moves 8 meters in the first three seconds. None of the given options match this calculation.

Figure 12: The example shows a representative response of the subcategory of responses with incorrect answers, that display incorrect visual interpretation (marked in bold), but correct reasoning.

**Correct vision, correct reasoning**

This subcategory contains very few responses, most of them from item 24 shown in the example below. Even if the selection of wrong answers could be seen as implying incorrect reasoning, we opted to code it as correct when the reasoning provided before the final answer was consistent with the visual interpretation of the item.

24. The graph to the right represents the velocity of an object´s motion. Which sentence is the best interpretation?

A) The object is moving with a constant acceleration.
B) The object is moving with a uniformly decreasing acceleration.
C) The position of the object decreases uniformly
D) The object is moving at a constant velocity.
E) The object does not move.

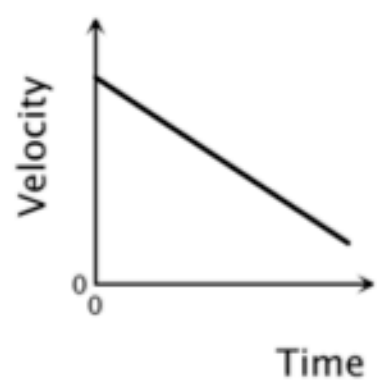

The graph shows that the velocity of the object is decreasing linearly with time. This means that the object is decelerating or slowing down at a constant rate.

Thus, the best interpretation is:
B) The object is moving with a uniformly decreasing acceleration.

Figure 13: The example shows one of the few cases displaying correct visual interpretation and correct reasoning (up until the final answer). It is difficult to fully understand why this answer was chosen, but taking into account the stochastic mechanisms underlying LLM-based models like ChatGPT, answers inconsistent with the reasoning are not a completely unexpected occurrence.

**Incorrect vision, incorrect reasoning**

In this category, most answers' incorrectness can be tracked down to faulty vision (or cases where visual interpretation could not be inferred directly), while they simultaneously also contain some mistakes in reasoning, incomplete reasoning or completely absent reasoning (common in item 8). The example shown below shows major faults in the visual interpretation of the graph, as well as some errors in physics reasoning. Namely, the statement that a slope cannot be determined for a single point on the graph is incorrect. The reason that the slope cannot be determined in these particular points is because of the discontinuity of the derivative, not the mere fact that they are points.

2. The following figure shows the velocity versus time graph of an object. Which of the following options corresponds to the case when its acceleration is the most negative?

A) V to X
B) T to V
C) V
D) X
E) X to Z

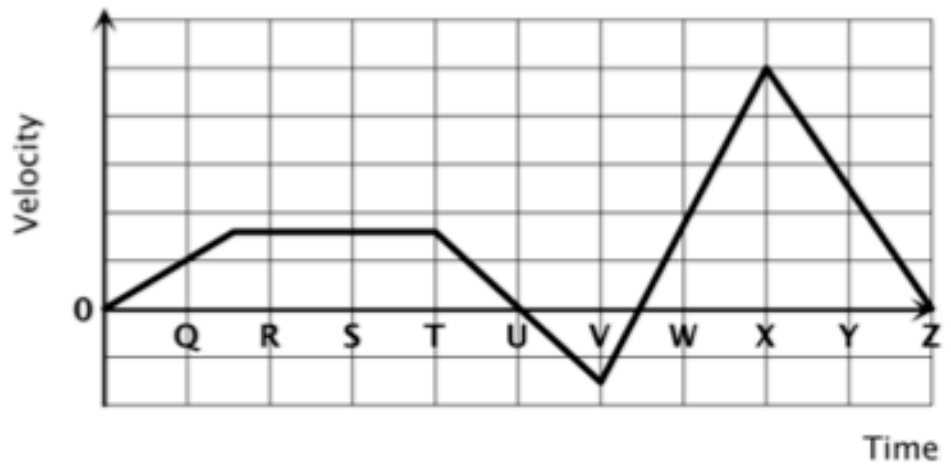

Acceleration is represented by the slope (or steepness) of a velocity versus time graph. The steeper the graph, the greater the magnitude of the acceleration. When the graph slopes downwards (negative slope), the acceleration is negative.

Let's analyze the given options:
A) **V to X: The graph has a steep negative slope**, indicating negative acceleration.
B) **T to V: The graph is horizontal**, indicating zero acceleration.
C)**V: This is a point on the graph, not a segment, so it doesn't represent a change in velocity over time.**
D) **X: Similar to option C, this is a point on the graph.**
E) **X to Z: The graph has a positive slope, indicating positive acceleration.**

Among the given options, the steepest negative slope (most negative acceleration) is from V to X.

Therefore, the correct answer is A) V to X.

Figure 14: The example shows a case of a response that contains both incorrect visual interpretation and mistakes in reasoning (marked in bold).

**Correct vision, incorrect reasoning**

This is when the reasoning was incorrect or not given, despite the vision being correct.

17 . The graph to the right shows the velocity of an object moving in a straight line. Which sentence is the best interpretation?

A) The object is moving with a uniformly increasing position.

B) The object´s position is constant.

C) The object is moving with a uniformly increasing acceleration.

D) The object is moving with a constant, non-zero acceleration.

E) The object is moving with a uniformly increasing velocity.

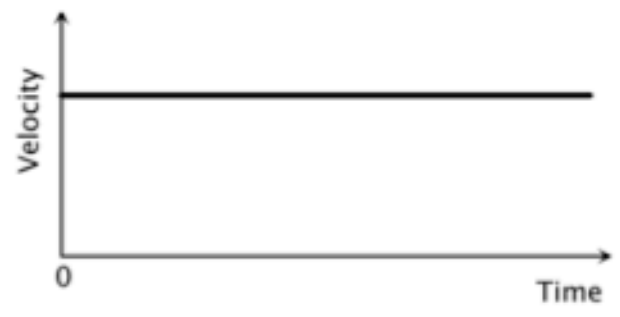

The graph shows that the velocity of the object is constant over time (i.e., the velocity does not change as time progresses).

Given the options:

A) Incorrect. Uniformly increasing position would imply a constant velocity, but not the other way around.

B) Incorrect. A constant position would mean the velocity is zero.

C) Incorrect. Uniformly increasing acceleration would mean the velocity graph is a curve that's getting steeper.

D) Correct. **The object is moving with a constant, non-zero acceleration because the velocity is constant and non-zero over time.**

E) Incorrect. Uniformly increasing velocity would mean the velocity graph is a straight line with a positive slope.

So, the best interpretation is:

D) The object is moving with a constant, non-zero acceleration.

Figure 15: The example shows a case of a response that contains a correct visual interpretation but mistakes in reasoning (marked in bold).